# Controller selection in a Wireless Mesh SDN under network partitioning and merging scenarios

Stefano Salsano, Giuseppe Siracusano, Andrea Detti, Claudio Pisa, Pier Luigi Ventre, Nicola Blefari-Melazzi
CNIT / University of Rome Tor Vergata

*Abstract* - **In this paper we consider a Wireless Mesh Network (WMN) integrating SDN principles. The Wireless Mesh Routers (WMR) are OpenFlow capable switches that can be controlled by SDN controllers, according to the wmSDN (wireless mesh SDN) architecture that we have introduced in a previous work. We consider the issue of controller selection in a scenario with intermittent connectivity. We assume that over time a single WMN can become split in two or more partitions and that separate partitions can merge into a larger one. We assume that a set of SDN controllers can potentially take control of the WMRs. At a given time only one controller should be the master of a WMR and it should be the most appropriate one according to some metric. We argue that the state of the art solutions for "master election" among distributed controllers are not suitable in a mesh networking environment, as they could easily be affected by inconsistencies. We envisage a "master selection" approach which is under the control of each WMR, and guarantees that at a given time only one controller will be master of a WMR. We designed a specific master selection procedure which is very simple in terms of the control logic to be executed in the WMR. We have implemented the proposed solution and deployed it over a network emulator (CORE) and over the combination of two physical wireless testbeds (NITOS and w-iLab.t).**

## I. Introduction

SDN systems have an increasing diffusion among networks operators due to, at least, three reasons. First, the presence of a centralized network controller eases the programming of network functions with respect to a distributed approach. Second, there is an open and standardized interface (e.g. OpenFlow) to deploy rules on network elements of the data plane and this allows an operator to easily integrate in its network of data and control devices of different vendors, without being tied to proprietary solutions. Third, removing the control complexity from the network elements will lead to their commoditization and to large cost reductions.

Among these three aspects, the centralization of control gives rise to some resiliency concerns, especially in environments where the network can be partitioned and the controller could become not reachable for some network devices. When network partitioning is the norm rather than an exception, like in MANET or DTN, fully decentralized solutions like IP routing seems to be most suited. However when partition rate is low and somewhat predictable, partially-decentralized solutions, e.g. based on controller replication, could be effective to maintain SDN operativeness. These conditions may occur in Wireless Mesh Networks (WMNs), like Community Networks (e.g. [1][2][3][4]), in which some parts of the network are inter-connected by long links that may temporary fail. Deploying a replica controller in these predictable network parts makes SDN operations resilient to the partition event. For instance, in Figure 1 the breakage of the link between Wireless Mesh Routers (WMRs) A and B partitions the WMN in two parts. However, both network parts include a controller and thus SDN/OpenFlow switches could continue to properly operate.

The distribution of SDN controller is already considered in the literature (e.g. [5][6][7]). It requires: first, a synchronization layer which guarantees consistency of data plane rules among the controllers and exposes a logical single controller to the programmer; second, a mechanism to assign a single master controller to each network switch.

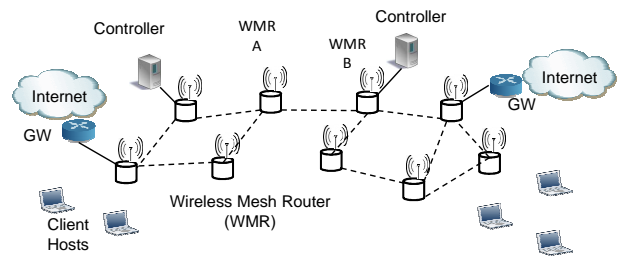

Figure 1 Reference Network scenario

Regarding the master assignment problem an explored approach is a kind of *master-election* [5], in which the controllers communicate each other to elect which is the master controller of a given switch and then inform the switch with an OpenFlow *Role request* message. This scheme assumes that communications among controllers are relatively reliable and is not meant to cope with partitioning issues but rather to provide load balancing functionality.

In this paper we focus the master assignment problem and propose a naïve approach of *master-selection*, rather than election. Each switch has a priority list of controllers. A switch autonomously selects as master the controller that can be contacted and has the highest priority. For load balancing purpose the switches could have a different priority list and the list could preconfigured or injected in the switches by the control plane. Controllers do not need to communicate each other as in the master-election approach.

A drawback of this approach is the deployment of additional control logic in the switch. This may be a simple algorithm, which periodically polls possible controllers. We have implemented the master-selection logic in a local process of a WMR, which locally configures the master controller of the switch. As a further step, the control logic could be implemented in the switch software or better in a "switchlet container" which should offer the notification of events like "controller connected/disconnected".

As a proof of concept, we verified the effectiveness of our master-selection implementation in both physical wireless network testbed environments ([10][11]) and in a single machine emulation platform based on CORE [14].

## II. SDN IN A WIRELESS MESH NETWORK: wmSDN

Our reference network scenario is shown in Figure 1. A WMN is composed of Wireless Mesh Routers (WMRs) which provide connectivity to a set of Access Networks (either offering a wired or wireless interface to client hosts). A subset of the WMRs operate as Gateways and provide connectivity towards the Internet. A set of OpenFlow controllers can operate in the wireless mesh each controller can be connected to the WMRs through wireless or wired connections. The scenario and the basic concepts described in this section have been proposed in [8] and are reported here for clarity. We refer to this architecture as *wmSDN* (wireless mesh SDN).

In wmSDN we can identify a control and data network. The control network is used for the exchange of routing information among the WMRs and for the communication between the WMRs and the SDN controllers. The data network is used by the Client hosts to communicate towards the Internet and among each other. We assume that range of IP addresses will be allocated for the control network (typically in the private IP address space) and each controller and mesh node will be statically given its IP address. Within the whole control network, each controller is uniquely identified by its IP address in the control network range. Under these assumptions the control plane connectivity can be built by using OLSR routing protocol. The data traffic will use a different set of IP subnets. For instance, the subnet 10.0.0.0/16 can be used for control traffic, while other subnets are used for data traffic (e.g. 192.168.x.0/24, always considering private IP addresses in these examples) and assigned to the Access Networks. The controllers and the WMR wireless interfaces over the WMN use addresses of the control subnet, while the interfaces towards the Access Networks gets an IP address belonging to the other subnets. The data traffic subnets are announced in OLSR as "HNA network" (HNA stands for Host and Network Association).

The proposed approach foresees to use an IP ad hoc routing protocol (OLSR) among the nodes of the mesh to establish a basic IP connectivity (see Figure 2). Such connectivity will constitute the control plane and will support all controller-to-switch OpenFlow messages as well as controller-to-controller messages in case they are needed to coordinate the SDN operations. The use of OLSR ensures the proper reaction to changing topology events, like addition/removals of mesh nodes and wireless links among them. To distribute the topology information of the data plane, the IP subnets of the Access Networks are advertised by WMRs and gateway WMRs using OLSR Host and Network Association (HNA) messages. The IP addresses of the controllers are also advertised using HNA messages with /32 mask. Moreover, gateway WMRs may also advertise the default route 0.0.0.0/0. With this approach, each WMR node knows the full network topology. The controllers inquiry the connected WMR to learn this topology information, which is fundamental to implement traffic engineering logic for data traffic. This approach is different from the traditional OpenFlow topology discover in wired layer 2 network, performed using LLDP messages

As for the wireless channels, we use a single SSID for both the control traffic and the data traffic, therefore we can classify it as an "in-band" control strategy from the OpenFlow protocol perspective.

In the wmSDN solution, the forwarding on the control plane will rely on the basic IP connectivity, while the data plane can use the IP connectivity or an "SDN based connectivity" in a flexible way. By SDN based connectivity we mean that the routing of packet flow is decided by the SDN controller and the forwarding within each node is based on the flow table rules installed using the OpenFlow protocol. In fact, each Wireless Mesh Router will also run a SDN capable OpenFlow switch.

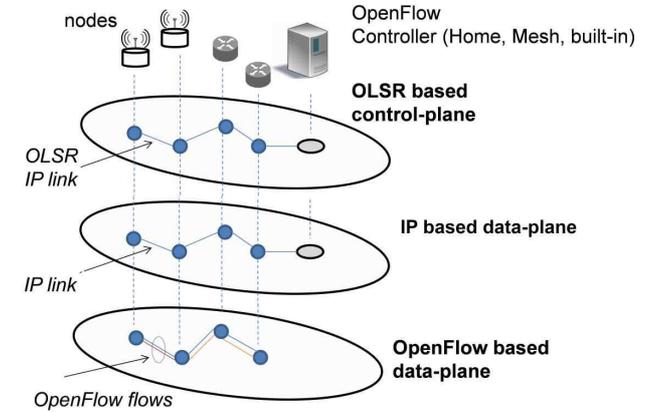

Figure 2 Control and data planes

## III. WMR NODE ARCHITECTURE

With respect to the wmSDN solution proposed in [8] we have redesigned the node architecture, by considering the issue of multiple controllers support and by improving the interaction between the switching and the routing components in the node. The high level node architecture is reported in Figure 3. Over the control plane, the OpenFlow switch can contact a set of controllers. A WMR node will also have a *built-in* module located in the switch itself for handling emergency services or, more in general, a network partition in which no network controllers are present. This built-in controller does not need to be a full compliant OpenFlow controller, rather it is a process that is able to inject OpenFlow rules in the local OpenFlow switch. We call this entity "EFTM" (Embedded Flow Table Manager).

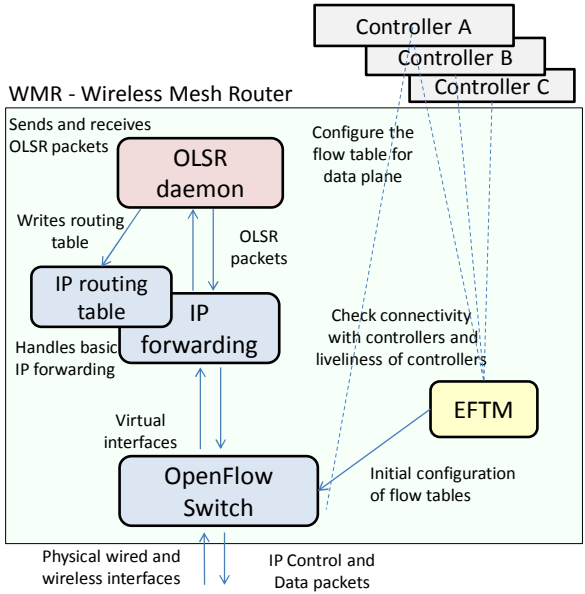

Figure 3 High level WMR node architecture

The OpenFlow controllers can be used to engineer the routing of data traffic, forcing an arbitrary subset of the traffic to follow a different route with respect to basic IP routing. The WMR nodes can connect to different controllers, supporting controller failures and dynamic topology modification, including network partitioning and joining. In emergency conditions, during which all OpenFlow controller fail or are unreachable, the basic IP routing is always available. The EFTM will also deal with the selection of the most appropriate controller, as detailed in section IV.

Figure 3 shows also the interplay between the OLSR protocol, the IP forwarding and the EFTM entity. The OpenFlow switch in the WMR is configured by the EFTM so that by default IP packets for the IP control subnet to which the WMR interfaces belong are handled by the IP forwarding modules. This way, the OLSR daemon can send and receive OLSR packets over the wireless interfaces. Once the IP routing tables are established with OSLR, the OpenFlow switch in the WMR interacts with the OpenFlow controllers that can configure the flow table for specific data plane flows.

A more detailed view of the WMR node architecture is reported in Figure 4. With respect to our original design reported in [8], we have integrated the approach of the OSHI (Open Source Hybrid IP/SDN) [9] framework for an Hybrid IP/SDN node. As shown in Figure 4, the OpenFlow capable switch is directly connected to a set of physical wireless interfaces belonging to the WMN (wlan0, wlan1). It could also be connected to wired interfaces (not shown in the figure) or to tunnel interfaces (tap9 in the figure). For each physical interface connected to the switch, a corresponding virtual internal interface is added to the OpenFlow switch. The physical interfaces do not have an IP address, virtual interfaces have IP addresses belonging to the control subnet. The IP routing and forwarding of the node operates using this set of virtual internal interfaces. Initially, a simple set of rules is configured in the switch so that the packets can flow from the physical interfaces to the virtual internal interfaces and vice versa. A packet that is sent by the IP layer in the WMR over a virtual interface crosses the OpenFlow switch and is sent out over the corresponding physical interface. An incoming packet arriving over a physical interface is forwarded by the OpenFlow switch to the corresponding virtual internal interface. The OLSR routing protocol runs using the virtual internal interfaces and learns the topology of the external links among the WMRs. As shown in Figure 4, additional interfaces can be directly visible to the IP forwarding/routing, without crossing the OpenFlow switch: ethY and wlanZ are the interfaces to the Access Networks, ethX is a wired interface connected toward the Internet for WMR gateways. The OLSR routing daemon is configured to work on the correct set of interfaces (e.g. the virtual internal interfaces), therefore it will not operate on the access networks nor on the interface towards Internet.

We note that OSHI solution [9] for the interaction between an OpenFlow switching component and the IP forwarding and routing component is very general and modular. We could replace the OSPF routing protocol (Quagga) used in [9] with the OLSR routing protocol (OLSRd) with minimal configuration effort.

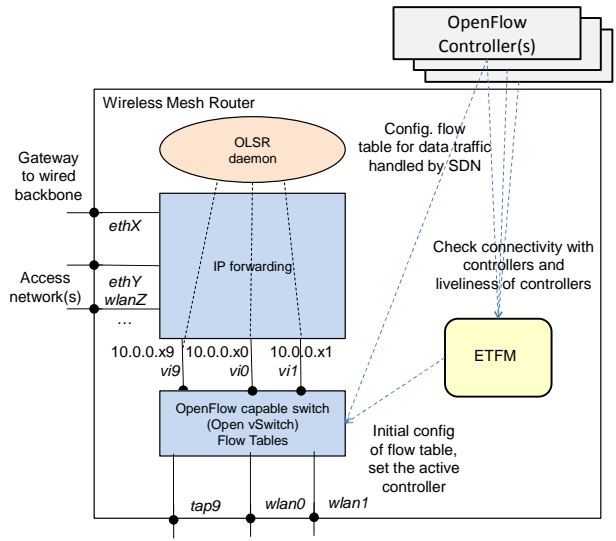

Figure 4 A more detailed view of WMR architecture

There will be two classes of packets/flows as seen by the OpenFlow switch in the WMR: i) packets/flows that are processed using regular IP routing/forwarding (*Basic* class); ii) packets/flows that will be handled by SDN (*SDN* class). For example, a possible approach is to include all traffic that belongs to the control-subnet in the Basic class and all traffic for IP destinations outside the control-subnet

(i.e. in the access networks or in the Internet) in the SDN class. The OpenFlow rules in the tables of the OpenFlow switches will be used to classify packets as belonging to Basic or SDN classes. A packet that belongs to the Basic class will be forwarded by the OpenFlow switch from the virtual internal interface to the physical interface (outgoing packets) or vice versa (incoming packets). A packet that belongs to the SDN class will need to find a matching entry in the flow tables of the switch or it will be forwarded to the OpenFlow controller. For this type of traffic within the Wireless Mesh Network a matching entry will specify the outgoing interface and will set as destination MAC address the next hop MAC address and as source MAC address the MAC address of the outgoing interface. In this way the OpenFlow switch will emulate the behavior of a OLSR router in forwarding the packet, but the outgoing interface can be set arbitrarily by the controller without following the routing chosen by OLSR.

The EFTM entity in the WMR will continuously check if the WMR is connected to an active controller. In case of controller failures (e.g. due to hardware or communication issue) the ETFM will trigger the start of an "emergency condition". In this state the EFTM can choose to clear all rules set by the controller, so that the node will only operate at IP level with the routing enforced by OLSR. In these conditions the EFTM could enforce some policies to handle data traffic. The most restrictive policy will be to allow only control traffic, i.e. directed towards the IP addresses of the control subnet, the most liberal policy will be to allow traffic towards all destinations (including all access networks all Internet destinations that are routed towards the default gateways advertised by OLSR), specific policies can be configured to selectively control which traffic has to be forwarded.

### IV. CONTROLLER SELECTION ARCHITECTURE

In a SDN scenario with multiple concurrent controllers, two important issues needs to be addressed: i) the different controllers need to share a common view of topology and of the network events that are relevant to take decisions in the controller layer; ii) the controllers need to synchronize about which controller is master for each switch, performing the so called "master election" procedure.

As for the common view of topology and events, we assume that the OLSR topology distribution mechanism is exploited by OpenFlow controllers. The controllers will learn the topology and will receive topology updates using OLSR. Considering the issue of controller discovery by WMRs, in our experiment the controllers are assigned IP addresses in a specific range. Therefore the WMR dynamically learns the existence of a controller when receiving OLSR messages advertising an IP address in the controller range. For further study we could address extensions to OLSR protocol to explicitly identify the controller addresses by tagging them in some way or transporting them in separate messages.

As for the "master election" procedure, in our scenarios it needs to be repeated each time that a portion of network become partitioned or when different partitions are joined together in a larger partition. In a traditional OpenFlow environment, it is assumed that communications among controllers are relatively reliable. Therefore the master election procedure can be executed with information exchange among controllers that cooperatively choose a master to take control of a given switch. Then the controllers send "role request" messages that are able to change controllers status, enforcing for example one "master" and a set of "slave" controllers for a given OpenFlow switch. An example in this line of reasoning can be found in [5]. Considering the requirements of a wireless mesh environment that includes topology changes and links unreliability, there is the risk that a distributed master election procedure produces inconsistent results. For transient periods, controllers could be connected with a WMR but could not be able to communicate each other. Under such circumstances, both controllers would believe they are in charge of controlling the WMR and would try to become "master" using the role request messages.

For this reason we designed a procedure in which the WMR itself is in charge of selecting the most appropriate controller given the connectivity status of the network. We note that WMRs and controllers have the same information about the status of the network (excluding transient conditions), because they share the OLSR vision of the topology. In particular, the WMRs are directly involved in the OLSR topology dissemination while the controller extracts the topology information from a nearby WMR. Therefore, from the topology discovery point of view the WMR acquires topology information even before the controller. Moreover, a WMR can directly check the connectivity with potential controllers trying to establish TCP connections towards them (or monitoring the liveliness of established TCP connections). In the designed procedure a WMR connects only toward a single controller at a given time. This is different from the classical approach where a switch connects in parallel with several controllers. The procedure is performed in the WMR with the help of the EFTM (External Flow Table Manager) that we have introduced for handling the flow tables. The EFTM is in charge to perform the master selection procedure and will instruct the switch to connect to the selected controller at a given time. We call the proposed mechanism as "master selection" rather than master election: it is directly the WMR node that monitors changes in the network topology (split/merging of mesh network, each such change can make unavailable/available a given controller). Following a network topology change, a WMR node takes into account the available (reachable) controllers, selects the best one (the highest in the hierarchy) and setups an OpenFlow control connection with it. From the implementation point of view, a sort of "hard" handover of the controller is performed by the WMR. The EFTM entity instructs the OpenFlow switch running in our WMR node (Open

vSwitch) to disconnect from the previous controller and to connect to the new one. The existing rules in the flow table are not changed, therefore the logic in the newly connected controller can decide if to delete all rules (which is the most reasonable choice in most of the scenarios) or if to leave existing rules active and rely on rule expiration.

We note that our mechanism is conceived to work for events of topology changes (network merging/joining) that operate in the time scale of tens of seconds, it may become critical if we want to manage such events in the time scale of few seconds (hysteresis timers can be added to the solution to protect from too frequent changes).

Performing the master selection on the WMR side has some advantages in our scenario. The first advantage is that each OpenFlow switch will be connected with a single controller at a time, and no conflicting rules can be injected. The second advantage is that a coordination mechanism among controllers is not needed, each controller can operate on its own.

The proposed architecture may support different algorithms for the controller selection by the WMRs, which can take into account static configuration information and/or dynamic information pushed into the WMRs by the controllers or gathered by the WMRs. We defined and implemented a rather simple approach with the purpose to demonstrate the effectiveness of WMR based master selection with practical experiments and measurements. In this solution, all controllers are organized in a hierarchy with strict ordering. The controller with the highest priority among those reachable by a WMR will be selected. Note that with this solution we focus on the problem of partitioning/merging of the WMN, while at the moment we are deliberately not focusing on load sharing issues. In fact, in our solution when the network is not partitioned, all WMRs will select as their master controller the highest controller of the hierarchy, which will become the controller of the whole network.

This solution is implemented by associating the priority to all controllers based on their IP address. Therefore we assign the IP addresses to the controllers so that the desired hierarchy is enforced. The available controllers are announced by OLSR as HNA (Host Network Association) and the WMR distinguishes the controllers assuming that their IP address will belong to a particular range. This simple solution does not require any enhancement to OLSR, more sophisticated solutions can be adopted extending OLSR so that the existence and priority of controllers can be explicitly advertised in OLSR announcements.

V. IMPLEMENTATION AND PERFORMANCE EVALUATION

We have implemented the proposed solution and deployed it in two environments: i) a distributed experimental setup over physical wireless nodes in NITOS and w-iLab.t testbeds, interconnected with Ethernet over UDP tunnels across PlanetLab Europe (see [12]); ii) a single machine environment using the CORE emulator [14]. In this section we report performance evaluation results obtained using the single machine emulation. The experiments are performed on a CORE emulator setup that integrates ns-3 for the simulation of wifi MAC protocols. We used Pox [13] as SDN controller. All the developed software has been released as Open Source and is available at [15] (to ease the replication of the experiments, a VirtualBox image with the development and test environment can be downloaded).

Figure 5 shows the emulated network used in our experiments, composed of 6 WMRs, two controllers, 3 wired access networks interconnected through the WMN. The link between wmr2 and wm3 is the "critical" one: we partition /merge the network by removing /adding this link.

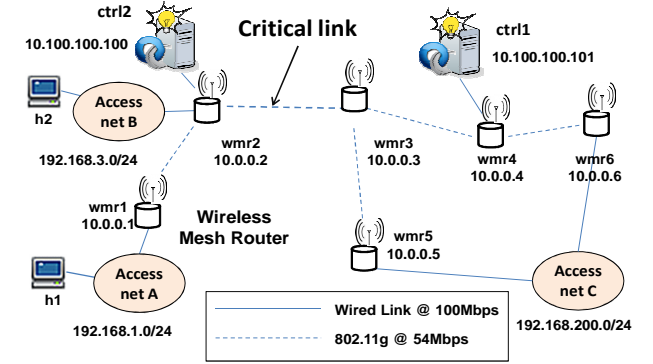

Figure 5 Emulated network

A. *Network merging experiment*

In this experiment we evaluate the time needed for the WMRs to connect to a higher priority controller after the merging of two network partitions. In the initial state the network is partitioned as the link between wmr2 and wmr3 is down (this is realized by means of some dropping rules in the flow tables of wmr2): wmr1 and wmr2 are connected to controller ctrl2. The dropping rules are then eliminated at the start of the experiment, allowing the OLSR protocol to discover the link between wmr2 and wmr3 and distribute the new topology to all WMRs. We measure the time needed for all WMRs (wmr1 and wmr2 in this case) to switch to controller ctrl1 (which has a higher priority). As shown in Figure 6, this time is decomposed in two phases, network connectivity and master selection. The former one considers the time needed for the routing protocol to setup the IP routes in all WMRs taking into account the merged network topology. We measure it by trying to send ping requests from h1 to ctrl1, the first ping reply received by h1 corresponds to the network connectivity interval. In our experiments it averages to 15 seconds. This is consistent with the OLSR routing protocol mechanisms, as three "Hello" messages needs to be received in order to declare a link up and the default interval for sending OLSR Hello messages is 5 seconds. By increasing the sending rate of OLSR messages it is possible to reduce the network connectivity time, at the price of increasing the OLSR processing and the link occupation overheads. Starting from this time instant we measure the interval needed for the two WMRs to disconnect from ctrl1 and connect to ctrls. In our implementation the EFTM periodically tries to establish a connection with all controllers that have been discovered, starting from the highest priority one. The polling period is 3 seconds, in the experiment we measured an averages of more than 1.5 seconds for the latest connected WMR, which is consistent with the expectations.

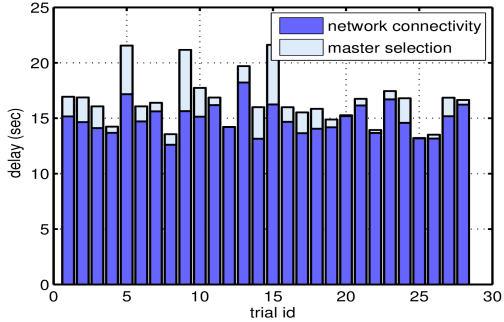

Figure 6 Network merging experiment

*B. Network partitioning experiments*

In this second set of experiments we consider the partitioning of the network: starting from the network topology shown in Figure 5, we disconnect the link between wmr2 and wmr3 (by means of some dropping rules in the flow tables of wmr2). In Figure 7 we report the evaluation of the time needed by wmr1 and wmr2 to disconnect from ctrl1 and connect to ctrl2 (the latest connection time is shown. In this case the WMRs does not rely on OLSR to discover that a controller is not reachable, as it would require more than 15 seconds considering the default OLSR configuration (3 Hello intervals of 5 s needed to declare the link down). The ETFM periodic controller polling procedure considers a 2 seconds timeout before declaring that a controller is down. With this procedure, an average master selection delay of 5.5 seconds is measured.

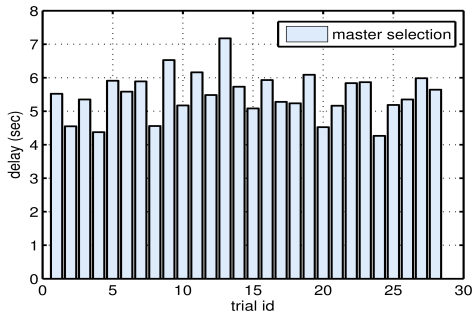

Figure 7 Network partitioning experiment: master selection delay

In the network partitioning scenario, we also measured the file transfer throughput between h1 in Access net A and h2 in Access net B (Figure 5). We assume that an SDN based path (initially setup by ctrl1) is used to carry traffic between h1 and h2. Then the network is partitioned (at time t1) and the WMRs connect to ctrl2. The new controller removes all dynamic flow tables, the subsequent IP packets arriving at the WMRs when the flow tables are empty will generate "packet-in" messages towards ctrl2 that will need to setup the correct route. As shown in Figure 8, this corresponds to a drop of throughput that is soon recovered after that the flow tables have been setup again.

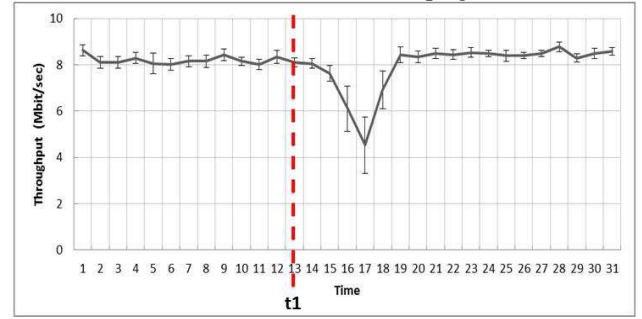

Figure 8 Network partitioning experiment: throughput of a TCP file transfer during controller handoff

## VI. CONCLUSIONS

In this paper we analyzed the issue of master controller selection in a wireless mesh SDN (wmSDN) scenario. Considering the peculiarity of this environment we have designed a solution in which the Wireless Mesh Routers (WMRs) are in charge to select their controller rather than being slave of a distributed controller election procedure. The proposed solution has been implemented and deployed in single-machine and in physical wireless testbed. Performance experiment in the single-machine deployment have been reported, showing results that are fully in line with the expectations.

## ACKNOWLEDGEMENTS

This work was partly funded by the EU in the context of the OpenLab and Confine projects.